\newif\ifjournal\journalfalse

\ifjournal
\documentclass[12pt,preprint]{aastex}
\else
\documentclass[iop]{emulateapj}
\journalinfo{The Astrophysical Journal, 712:951--956, 2010 April 1}
\submitted{Submitted 2009 November 12; accepted 2010 February 13; published 2010 March 8}
\fi
\shorttitle{Anomalous Boost in Relativistic Jet}
\shortauthors{Zenitani, Hesse, \& Klimas}
\def\jcp{{\itshape J.~Comput.~Phys.}}

\usepackage{graphicx}
\usepackage{bm}
\renewcommand{\vec}[1]{\boldsymbol{#1}}

\newcommand{\grad}{\nabla}
\renewcommand{\div}{\nabla \cdot}

\begin{document}

\title{Scaling of the Anomalous Boost in Relativistic Jet Boundary Layer}

\author{Seiji Zenitani, Michael Hesse, and Alex Klimas}
\affil{
NASA Goddard Space Flight Center, Greenbelt, MD 20771, USA; {Seiji.Zenitani-1@nasa.gov}
}

\begin{abstract}
We investigate the one-dimensional interaction of
a relativistic jet and an external medium.
Relativistic magnetohydrodynamic simulations show
an anomalous boost of the jet fluid in the boundary layer,
as previously reported. 
We describe the boost mechanism
using an ideal relativistic fluid and magnetohydrodynamic theory.
The kinetic model is also examined for further understanding.
Simple scaling laws for the maximum Lorentz factor
are derived, and verified by the simulations.
\end{abstract}

\maketitle

\section{Introduction}

Relativistic jets are considered
in various contexts in high-energy astrophysics,
such as active galactic nuclei (AGNs) \citep{up95,ferrari98},
microquasars \citep{mirabel99},
and potentially gamma-ray bursts (GRBs)
\citep{piran04,mes06}. 
The interaction between fast moving jets
(the relevant Lorentz factors are
$\gamma_{jet} \sim 10$--$20$ in AGNs and
$\gamma_{jet} \gtrsim 10^2$ in GRBs)
and
the surrounding medium is very important
to understand global dynamics of the jet system,
because it is related to the mass, momentum, and energy transport
across the boundary layers. 
In this context,
development of velocity shear instabilities has been of interest
(\citet{tur76,bp76,ferrari80,birk91,bodo04,osm08} and references therein). 
Moreover, a relativistic jet-medium boundary is
a potential site of high energy particle acceleration
as well \citep{ost00,so02}. 

Recently, it has been reported that
the jet-medium interaction is more complex than thought
even in the simplest one-dimensional (1D) case.
Raising a Riemann problem of relativistic hydrodynamics (RHD),
\citet{aloy06} showed that
the tangential hydrodynamic velocity and
the relevant Lorentz factor ($\gamma_{BL}$)
in the boundary layer
are anomalously accelerated ($\gamma_{BL}>\gamma_{jet}$)
when the jet is over-pressured. 
\citet{mizuno08} studied relativistic magnetohydrodynamic (RMHD) effect,
and reported that the perpendicular magnetic field
enhances the boost effect.
\citet{kom09b} discussed a similar tangential boost
in their RMHD simulation of the collapser jet. 
Such anomalous boost effect may be responsible for
increasing the jet's Lorentz factor \citep{aloy06,mizuno08}
and for modulating the radiative signature of the jet \citep{aloy08}.
However, its physical mechanism remains unclear,
and therefore no quantitative analysis has been performed.

In this paper,
we study the mechanism of the anomalous boost
by using RHD/RMHD simulations and an analytic theory.
In Section 2, we describe the problem setup.
In Section 3, we present the simulation results.
In Section 4, we construct an RHD/RMHD theory of the problem.
In Section 5, we additionally discuss kinetic aspects.
The last section Section 6 contains discussions and summary.

\section{Problem setup}

\begin{figure}[htbp]
\begin{center}
\ifjournal
\plotone{f1.eps}
\else
\plotone{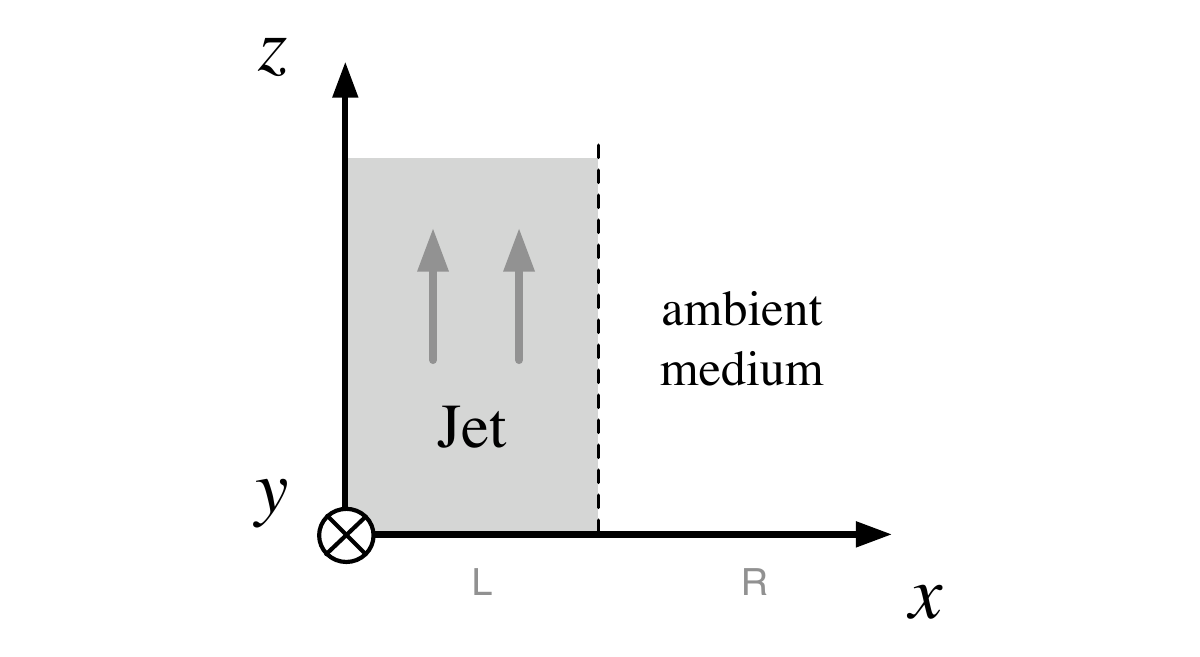}
\fi
\caption{
Our jet geometry.
We consider the jet in the left side ($L$) and
the ambient medium in the right side ($R$).
\label{fig:jet}}
\end{center}
\end{figure}

Following earlier works \citep{aloy06,mizuno08},
we study a 1D Riemann problem
in a jet-like configuration,
which is schematically illustrated in Figure \ref{fig:jet}. 
A jet travels upward in the $+z$-direction
in a stationary ambient medium. 
An interaction between the jet and the medium is
considered in the $x$-direction, and
we assume $\partial_y=\partial_z=0$.
Initially they are separated by a discontinuity and
we study the time evolution of this 1D system.

We employ the following ideal RMHD equations \citep{anile89}.
For convenience we set $c=1$ and
employ Lorentz--Heaviside units such that all $({4\pi})^{1/2}$ factors disappear.
\begin{eqnarray}
&& \partial_t(\gamma \rho) + \div (\gamma \rho \vec{v}) = 0
\label{eq:rmhda} \\
&& \partial_t \vec{m} + \div ( \gamma^2 w_t \vec{v}\vec{v} - \vec{b}\vec{b} + p_t \vec{I} ) = 0 
\label{eq:rmhdb} \\
&& \partial_t \mathcal{E} + \div \vec{m} = 0
\label{eq:rmhdc} \\
&& \partial_t \vec{B} + \div ( \vec{v}\vec{B} - \vec{B}\vec{v} ) = 0
\label{eq:rmhdd}
\\
&& \vec{E} + \vec{v} \times \vec{B} = 0 
\label{eq:rmhde}
\end{eqnarray}
\begin{eqnarray}
& \left\{
\begin{array}{ccl}
\vec{m} &=& \gamma^2 w_t\vec{v} - b_0\vec{b}
= \gamma^2 \rho h \vec{v} + (\vec{E}\times\vec{B})
\\
\mathcal{E} &=& \gamma^2w_t - b_0 b_0 - p_t 
\\
\vec{b} &=&
({\vec{B}}/{\gamma}) + \gamma (\vec{v}\cdot\vec{B})\vec{v} \\
b_0 &=& \gamma(\vec{v}\cdot\vec{B}) \\
w_t &=& \rho h + b^2 = \rho+{\Gamma p_g}/({\Gamma-1}) + b^2 \\
p_t &=& p_g + \frac{1}{2} b^2 \\
p_g &=& \rho T
\end{array}
\right.
\end{eqnarray}
In the above equations,
$\gamma$ is the Lorentz factor,
$\rho$ is the proper mass density,
$\vec{v}$ is the velocity,
$\vec{m}$ is the momentum density,
$\mathcal{E}$ is the energy density,
$w_t$ is the total enthalpy,
$h$ is the specific enthalpy,
$p_t$ is the total pressure,
$p_g$ is the gas pressure,
$T$ is the gas temperature including the Boltzmann constant,
and $b^{\alpha} = ( b_0, \vec{b} )$ is the covariant magnetic field.
Note that
$b^2=b^{\alpha}b_{\alpha}=B^2/\gamma^2+(\vec{v}\cdot\vec{B})^2=(B^2-E^2)$
is a Lorentz invariant.
We use an equation of state
with a constant polytropic index of $\Gamma=4/3$.

We developed an RMHD code to numerically solve the problem.
We employ a relativistic HLLD scheme \citep{mig09,miyoshi05},
which considers multiple states
inside the Riemann fan
in order to resolve discontinuities better. 
We interpolate the spatial profile by a monotonized central limiter \citep{mc} 
and
solve the temporal evolution by the second order
total variation diminishing (TVD) Runge--Kutta method. 
Relativistic primitive variables are
recovered by \citet{mig07b}'s inversion scheme. 

The model parameters are presented in Table \ref{table}.
The subscripts $L$ and $R$ denote the properties in the two regions
($L$ for the left side or the jet, and
$R$ for the right side or the ambient medium).
The Lorentz factor of the jet is set to $\gamma_{jet}=7$.
We initially set $B_{x}=0$.
In our 1D configuration
this automatically means $B_x=0$ all the time.
This condition $B_{x}=0$ allows us to simplify the numerical scheme,
because a five wave HLLD problem is reduced to
a three-wave problem (see \citet{mig09}, Section 3.4.1).
The first model H1 has no magnetic fields (RHD).
The other two models contain magnetic fields inside the jet:
the jet-aligned magnetic field ($B_z$: model M1)
and the out-of-plane magnetic field ($B_y$: model M2). 
Importantly, the total pressure $p_{t,L}$ is set to the same.
These RMHD models are analogous to
the ``poloidal'' (M1) and ``toroidal'' (M2) cases in \citet{mizuno08}.
The spatial domain of $-0.2 \le x \le 0.2$ is resolved by 6400 grids.
All simulation results are checked by an analytic solver by \citet{giac06}.

\ifjournal
\begin{deluxetable}{l|rccccccccc|rcccccccc}
\rotate
\tablewidth{0pt}
\tabletypesize{\scriptsize}
\else
\begin{deluxetable*}{l|rccccccccc|rcccccccc}
\fi
\tablecaption{\label{table} List of Simulation Models}
\tablehead{
\colhead{Model}
& \multicolumn{10}{c|}{Left} & \multicolumn{9}{c}{Right}
\\
\colhead{}
& \colhead{$\rho_L$} & \colhead{$p_{g,L}$}
& \colhead{$v_{x,L}$} & \colhead{$v_{y,L}$} & \colhead{$v_{z,L}$} & \colhead{$\gamma_{jet}$}
& \colhead{$B_{x,L}$} & \colhead{$B_{y,L}$} & \colhead{$B_{z,L}$} & \colhead{$p_{t,L}$}
& \colhead{$\rho_R$} & \colhead{$p_{g,R}$}
& \colhead{$v_{x,R}$} & \colhead{$v_{y,R}$} & \colhead{$v_{z,R}$}
& \colhead{$B_{x,R}$} & \colhead{$B_{y,R}$} & \colhead{$B_{z,R}$} & \colhead{$p_{t,R}$}
}
\startdata
H1 (RHD) & 0.1 & 10 & 0 & 0 &0.99 & 7 & 0 & 0 & 0 & 10 &
1 & 1 & 0 & 0 & 0 & 0 & 0 & 0 & 1 \\
M1 (RMHD) & 0.1 & 2 & 0 & 0 &0.99 & 7 & 0 & 0 & 4 & 10 &
1 & 1 & 0 & 0 & 0 & 0 & 0 & 0 & 1
\\
M2 (RMHD) & 0.1 & 2 & 0 & 0 &0.99 & 7 & 0 & 28 & 0 & 10 &
1 & 1 & 0 & 0 & 0 & 0 & 0 & 0 & 1
\\
\enddata
\tablecomments{
Models and parameters for the Riemann problems.
The subscript $L$ denotes the jet (left side) properties and
$R$ for the ambient medium (right side).
In addition, two parameter surveys are performed
by changing $\rho_{L}=(10^{-2},10^{-3})$ and $p_{g,R}=(3, 0.3, 0.1)$.
}
\ifjournal
\end{deluxetable}
\else
\end{deluxetable*}
\fi

\section{Results}

\begin{figure}[bhtp]
\begin{center}
\ifjournal
\includegraphics[width={0.8\columnwidth},clip]{f2.eps}
\else
\includegraphics[width={\columnwidth},clip]{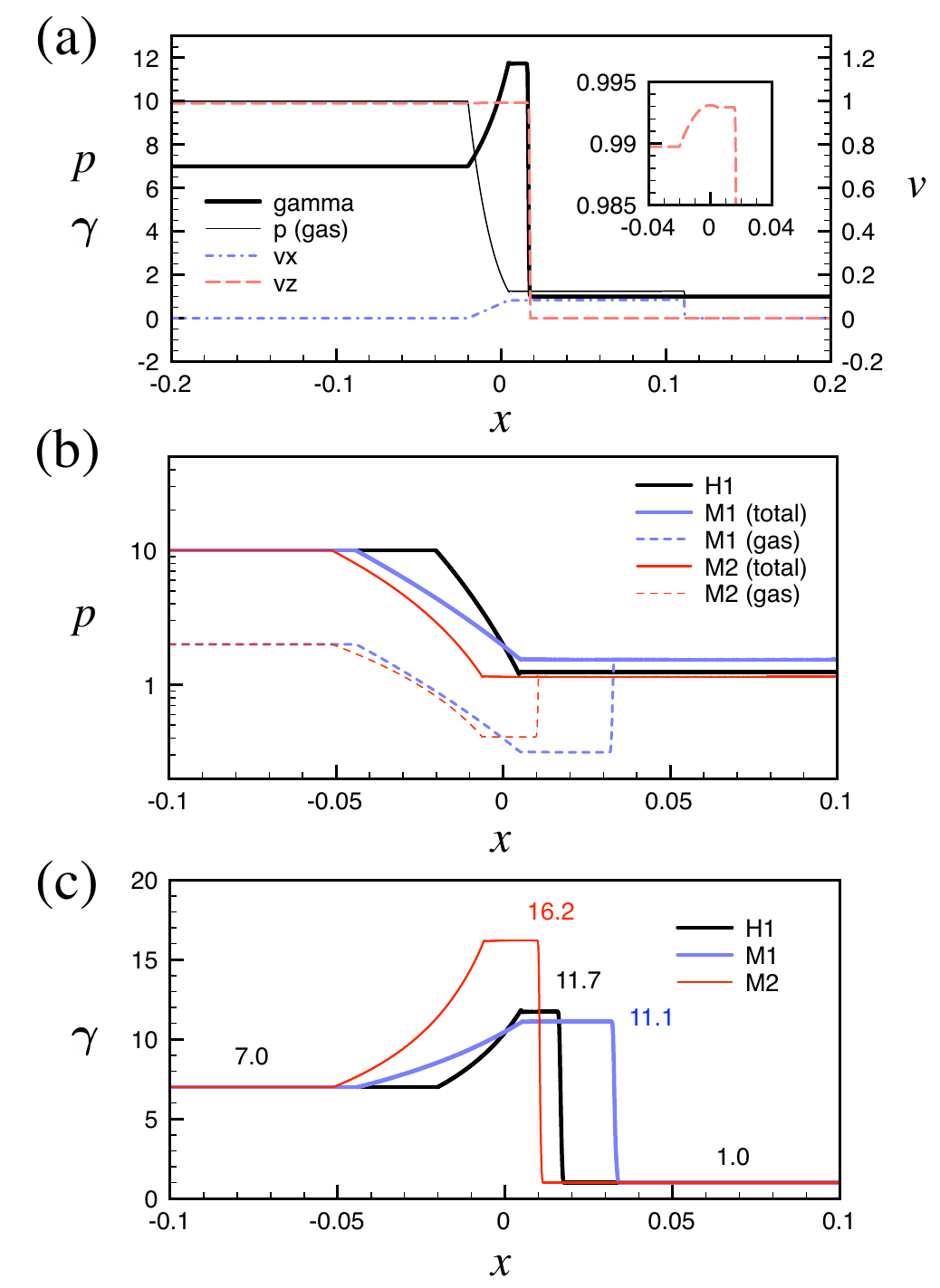}
\fi
\caption{
(Color online)
(a) Simulation result of model H1 at $t=0.2$.
The fluid Lorentz factor $\gamma$, the gas pressure $p_{g}$,
the normal velocity $v_x$, and the tangential velocity $v_z$ are presented.
The tangential velocity $v_z$ in the boosted region is also zoomed up
in the small box.
(b) The total pressure $p_t$ ({\itshape solid lines}) and
the gas pressure $p_g$  ({\itshape dashed lines}) 
in models H1 ({\itshape black}),
M1 ({\itshape red thin line}), and M2 ({\itshape blue thick line}) at $t=0.2$.
(c) The Lorentz factor $\gamma$ in three models at $t=0.2$.
The small numbers indicate the Lorentz factors in the relevant flat regions.
\label{fig:profile}}
\end{center}
\end{figure}

Shown in Figure \ref{fig:profile}{\itshape a} are
simulation results of the model H1 at $t=0.2$.
One can recognize a three-wave structure:
(1) a leftward rarefaction wave ($x \sim -0.02$),
(2) a contact discontinuity
that separates the jet and the ambient medium ($x \sim 0.02$),
and (3) a right-going forward shock ($x \sim 0.12$).
The system exhibits a self-similar evolution
as those waves propagate in time. 
Numerical errors are negligible,
thanks to the high resolution and the stable numerical scheme.
In the rarefaction region between (1) and (2),
the Lorentz factor of the fluid gradually increases from $\gamma_{jet} = 7$,
and then it reaches to the maximum ($\sim 11.7$)
at the left vicinity of the contact discontinuity. 
This is consistent with the anomalous boost
demonstrated in previous works.
Hereafter, we denote this boosted region as the ``boundary layer'' and
define the relevant Lorentz factor in the flat region $\gamma_{BL}$. 
The tangential velocity increases there,
as shown in the small box in Figure \ref{fig:profile}{\itshape a}. 

The RMHD models evolve similarly as the RHD model H1 does. 
Figure \ref{fig:profile}{\itshape b} compares
the pressure profiles of the three models,
and Figure \ref{fig:profile}{\itshape c} shows
the profiles of the Lorentz factor.
Since the jet contains the magnetic field in the RMHD cases,
the rarefaction wave fronts propagate faster than the RHD case,
because the Alfv\'{e}n speeds ($\sim c$ in the proper frames) are faster than
the sound speed ($c_s \sim {c}/{\sqrt{3}}$ in the proper frame).
One can also see the tangential discontinuities
between the jet and the ambient medium
($x \sim 0.03$ in model M1, $x \sim 0.01$ in M2),
where the magnetic pressure disappears and
the gas pressure suddenly increases to maintain the total pressure.
The anomalous boost similarly takes place
on the jet side of the those discontinuities. 
The forward shocks are just out of sight from figures in RMHD cases.
As reported by \citet{mizuno08},
the model M2 with a perpendicular magnetic field ($B_y$) exhibits
stronger boost ($\gamma_{BL} \sim 16.2$) than
the model M1 with a parallel magnetic field ($B_z$) ($\gamma_{BL} \sim 11.1$).

\section{Analytic theory}

\subsection{RHD theory}

In this section
we study the mechanics of the anomalous boost problem.
First we examine the RHD case.
Combining the momentum equation (Equation \ref{eq:rmhdb}) and
the energy equation (Equation \ref{eq:rmhdc}) \citep{sakai},
\begin{eqnarray*}
\partial_t (\gamma^2\rho h\vec{v})
+ \vec{v} \Big( \div ( {\gamma^2 \rho h \vec{v}} ) \Big)
+ {\gamma^2 \rho h} (\vec{v} \cdot \grad) \vec{v}
+ \grad p_g = 0
, \\
\vec{v} \partial_t (\gamma^2\rho h) - \vec{v} \partial_t p_g
+ \vec{v} \Big( \div ( {\gamma^2 \rho h \vec{v}} ) \Big) = 0
,
\end{eqnarray*}
we obtain
\begin{eqnarray}
\label{eq:dpdt}
\gamma^2 \rho h \frac{D \vec{v}}{Dt} = -\grad p_g - \vec{v} \frac{\partial p_g}{\partial t}.
\end{eqnarray}
Since $\partial_z=0$,
the anomalous boost obviously comes from the last term,
$\gamma^2 \rho h (D/Dt) v_z \sim - \partial_t p_g$. 
This term has no Newtonian counterpart,
it is certainly a relativistic effect. 
In usual contexts, the term slows down the fluid bulk acceleration
in the high-temperature regime ($p_g \gtrsim \rho$),
as if the relativistic pressure increases the inertia.
In this case, since the pressure decreases in the rarefaction region,
the force in the last term boosts the fluid in the $z$-direction,
until the fluid element reaches the constant-pressure region. 
We see that the term coverts excess
internal energy to the energy of the bulk motion.

Next, we arrange
the momentum equation (Equation \ref{eq:rmhdb}) in the following way.
\begin{eqnarray*}
\label{eq:p_force}
\gamma\rho (\partial_t + \vec{v} \cdot \grad) ( \gamma h \vec{v} ) +
\Big[ \partial_t ( \gamma\rho ) + \div ( \gamma \rho \vec{v} ) \Big]
\gamma h \vec{v}
= - \grad p_g .
\end{eqnarray*}
Using Equation \ref{eq:rmhda}, we obtain
\begin{eqnarray}
\label{eq:constP}
\gamma\rho \frac{D}{Dt} ( \gamma h v_z ) = 0.
\end{eqnarray}
Thus, the specific momentum
(the momentum density per the gas density in this frame)
remains constant {\itshape as it should be}.
This is because no external forces accelerate the fluid, and
because the ideal fluid assumption does not allow
momentum transport in its own frame.
We confirmed that
$\gamma h v_z$ is well conserved in both sides in the simulation.

In model H1,
the jet velocity is initially relativistic ($v_{z,L}\sim 1$),
and then we expect
\begin{eqnarray}
\label{eq:rh}
\gamma h \sim {\rm const.}
\end{eqnarray}
in the rarefaction region.
The behavior of Equation \ref{eq:rh} is controlled by
the gas temperature, $T=(p_g/\rho)$.
When the gas is cold ($T \ll 1$),
both the specific enthalpy $h \sim 1$ and
the Lorentz factor $\gamma$ remain constant;
no boost occurs.
When the gas is relativistically hot ($T \gg 1$),
$h \sim 4T$ becomes a function of $T$.
In this limit, we find
\begin{eqnarray}
\label{eq:rT}
\gamma T = \gamma ({p_g}/{\rho}) \sim {\rm const.}
\end{eqnarray}
We see that the Lorentz factor increases
when the relativistic temperature decreases.
Physically this is relevant to
the temporal decrease of the pressure (Equation \ref{eq:dpdt}).
Combining with the polytropic law (${p_g}{\rho^{-\Gamma}} = {\rm const.}$),
we obtain the following relations,
\begin{eqnarray}
\label{eq:rho}
\gamma \rho^{\Gamma-1} \sim {\rm const.} \\
\label{eq:p}
\gamma p_g^{(\Gamma-1)/\Gamma} \sim {\rm const.}
\end{eqnarray}

Using these relations, we can estimate the boosted Lorentz factor $\gamma_{BL}$.
Inside the rarefaction region,
the gas pressure decreases to that of the contact discontinuity ($p_{g,D}$).
Since $p_{g,D} \gtrsim p_{g,R}$,
we immediately obtain the upper bound of $\gamma_{BL}$,
\begin{eqnarray}
\label{eq:max}
\gamma_{BL} \sim \gamma_{jet} \Big(\frac{p_{g,L}}{p_{g,D}}\Big)^{(\Gamma-1)/\Gamma}
\lesssim \gamma_{jet} \Big(\frac{p_{g,L}}{p_{g,R}}\Big)^{1/4}
.
\end{eqnarray}
It is interesting to see that
$\gamma_{BL}$ is controlled by the external pressure $p_{t,R}$. 
The over-pressured jet pushes the discontinuity outward,
and the external pressure terminates the boost
by stopping the further development of the rarefaction structure. 
The external pressure does no mechanical work on the jet fluid. 

Note that the boost does not operate
when the jet-side pressure becomes nonrelativistic ($T \lesssim 1$).
We have another restriction from Equation \ref{eq:rT},
\begin{eqnarray}
\label{eq:max2}
\gamma_{BL} \ll \gamma_{jet} \Big(\frac{p_{g,L}}{\rho_{L}}\Big).
\end{eqnarray}
This will replace Equation \ref{eq:max},
when the external pressure is too low ($p_{g,R}\rightarrow 0$).

We also examine the energy equation.
Inside the over-pressured ($p_{g}\gg \rho$) and
relativistically-moving ($4\gamma_{jet}^2 \gg 1$) jet,
the fluid energy density is
\begin{eqnarray}
\label{eq:ene1}
\mathcal{E} = ( \gamma^2 w_t - p_{g} ) \sim (\gamma \rho) \gamma h .
\end{eqnarray}
Substituting Equation \ref{eq:ene1} into Equation \ref{eq:rmhdc},
we obtain the same condition as Equation \ref{eq:rh}:
\begin{eqnarray}
\label{eq:ene2}
\gamma\rho (\partial_t + \vec{v} \cdot \grad) ( \gamma h ) +
\Big[ \partial_t ( \gamma\rho ) + \div ( \gamma \rho \vec{v} ) \Big]
\gamma h
\nonumber \\
= \gamma\rho \frac{D}{Dt} ( \gamma h ) = 0 .
\end{eqnarray}
Equations \ref{eq:ene1} and \ref{eq:ene2} tell us that
a specific energy density (the energy density per the lab-frame gas density)
is conserved during the fluid convection.
This is because
the total energy flow ($\gamma^2w_t\vec{v}\sim 4\gamma^2p_g\vec{v}$) is much larger
than the work to expand the jet outward ($p_g\vec{v}$),
and because the ideal fluid contains no heat transfer in its proper frame.

\subsection{RMHD theory}

Let us consider the effect of the jet-aligned magnetic field,
$\vec{B}_L=(0,0,B_z)$.
After some algebra in Equations \ref{eq:rmhdb},
we find both the $z$-momentum and
the $xz$ component of the stress-energy tensor
are unchanged from hydrodynamic ones.
Therefore we can utilize Equations \ref{eq:constP} and \ref{eq:rh}. 
We further consider flux conservation,
\begin{eqnarray}
\frac{B_z}{\gamma \rho} = {\rm const.}
\end{eqnarray}
Combining this with Equation \ref{eq:rho}, we obtain
\begin{eqnarray}
\label{eq:B}
\gamma B_z^{({\Gamma-1})/({2-\Gamma})} \sim {\rm const.}
\end{eqnarray}
When $v_z \sim 1$ like the boosted rarefaction region,
the magnetic pressure approximates $\frac{1}{2}b^2 \sim \frac{1}{2}B^2_{z}$.
From Equations \ref{eq:p} and \ref{eq:B}, we construct
the pressure condition across the tangential discontinuity,
\begin{eqnarray}
p_{g,L} \Big( \frac{\gamma_{jet}}{\gamma_{BL}} \Big)^{\frac{\Gamma}{\Gamma-1}}
+ \frac{B^2_{z,L}}{2}  \Big( \frac{\gamma_{jet}}{\gamma_{BL}} \Big)^{\frac{2(2-\Gamma)}{\Gamma-1}}
\sim p_{t,D} \gtrsim p_{t,R}.
\end{eqnarray}
The power indexes are both $4$ when $\Gamma=4/3$.
Therefore we obtain a generalized upper bound,
\begin{eqnarray}
\label{eq:max_para}
\gamma_{BL} \lesssim
\gamma_{jet} \Big(\frac{p_{t,L}}{p_{t,R}}\Big)^{1/4}
.
\end{eqnarray}
Note that the total pressure $p_t$ replaces
the gas pressure $p_g$ in Equation \ref{eq:max}.

In the case of the perpendicular magnetic field, $\vec{B}_L=(0,B_y,0)$,
the initial choice of $v_y=0$ simplifies the equations (e.g., $b_0=b_x=0$),
because both $v_y$ and $B_z$ remain zero \citep{romero05}.
In this case, the boost comes from
the temporal decrease of the total pressure,
$\gamma^2 w_t (D/Dt) v_z \sim - \partial_t p_t$.
From Equations \ref{eq:rmhda} and \ref{eq:rmhdb},
we can similarly derive the conservation law,
\begin{eqnarray}
\label{eq:by_const}
\gamma\rho \frac{D}{Dt} \Big( \gamma \frac{w_t}{\rho} v_z \Big) &=&
\gamma\rho \frac{D}{Dt} \Big( \gamma ( h+ \frac{b^2}{\rho}) v_z \Big) = 0
.
\end{eqnarray}
For simplicity, we consider the magnetically dominated limit of
$b^2/\rho \gg h$ (or $b^2 \gg 4p_g$).
In the jet side ($v_z\sim 1$) we expect $\gamma b^2 / \rho \sim {\rm const.}$
Combining this with the flux conservation
\begin{eqnarray}
\label{eq:by}
\Big( \frac{B_y}{\gamma\rho} \Big)^2 =
\frac{b^2}{\rho^2} = {\rm const.}
,
\end{eqnarray}
we expect
\begin{eqnarray}
\gamma^2 b^2 \sim {\rm const.}
\end{eqnarray}
The condition across the discontinuity leads to an upper bound of $\gamma_{BL}$,
\begin{eqnarray}
p_{t,L} \Big( \frac{\gamma_{jet}}{\gamma_{BL}} \Big)^{2} \sim
\frac{b_L^2}{2}\Big( \frac{\gamma_{jet}}{\gamma_{BL}} \Big)^{2}
\sim p_{t,D} \gtrsim p_{t,R}
,\\
\label{eq:max_perp}
\gamma_{BL} \lesssim \gamma_{jet} \Big(\frac{p_{t,L}}{p_{t,R}}\Big)^{1/2}.
\end{eqnarray}
Furthermore, from the polytropic law and Equation \ref{eq:by},
we see that the magnetic pressure decays more rapidly than the gas pressure,
\begin{eqnarray}
b^2 \propto p_g^{2/\Gamma} \sim p_g^{3/2}.
\end{eqnarray}
Consequently, the system behaves similarly as the hydrodynamic case
once the gas contribution and the magnetic contribution become comparable.
Therefore, we usually expect
intermediate results between Equations \ref{eq:max_para} and \ref{eq:max_perp}.

Among the two RMHD cases,
the boost is more significant in the perpendicular case
than in the parallel case \citep{mizuno08}.
This is because more electromagnetic energy and momentum are available per a gas medium
--- the jet initially contains larger field energy $\frac{1}{2}(B^2+E^2)$ and
carries additional upward momentum
in a form of Poynting flux ($\vec{E}\times\vec{B}$). 
We also recall that the boost process is related to the pressure decrease,
and that the magnetic pressure preferably works in the perpendicular directions.

\begin{figure}[hbtp]
\begin{center}
\ifjournal
\includegraphics[width={\columnwidth},clip]{f3.eps}
\else
\includegraphics[width={\columnwidth},clip]{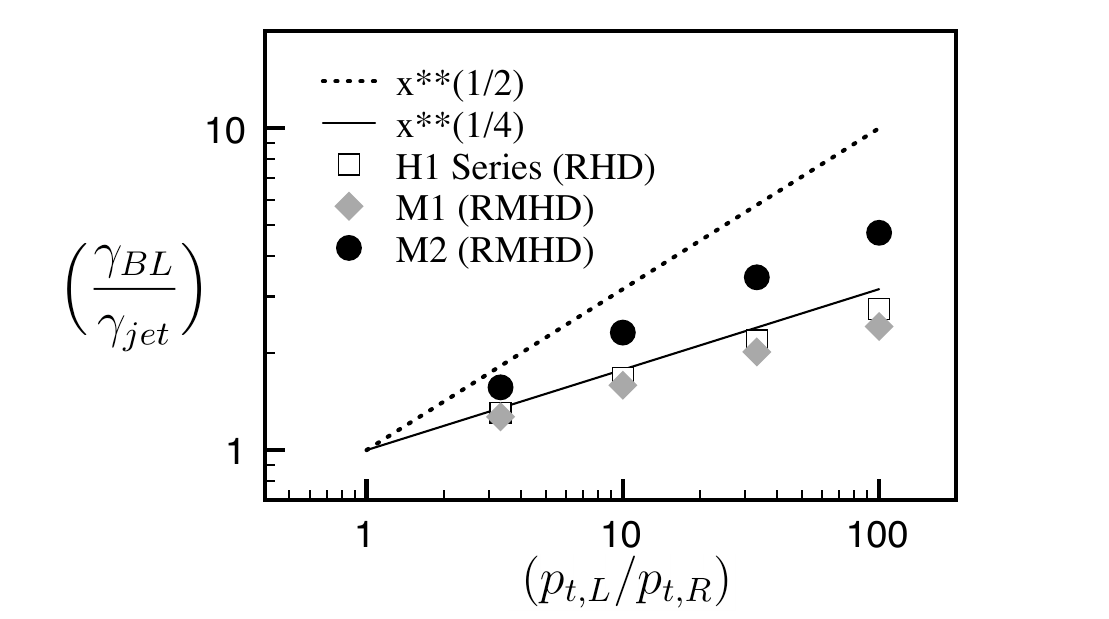}
\fi
\caption{
Anomalous boost ($\gamma_{BL}/\gamma_{jet}$)
as a function of the total pressure ($p_{t,L}/p_{t,R}$). 
Three models (H1, M1, and M2) are compared with the theories:
Equations \ref{eq:max} and \ref{eq:max_para} ({\itshape solid line}) and
Equation \ref{eq:max_perp} ({\itshape dotted line}).
\label{fig:max}}
\end{center}
\end{figure}

\subsection{Numerical Tests}

In order to verify the scaling theory,
we carry out series of parameter surveys,
by controlling the external pressure,
$p_{t,R}=p_{g,R}$ (Table \ref{table}).
Figure \ref{fig:max} shows
the boosted Lorentz factors ($\gamma_{BL}$)
in our RMHD simulations
as a function of ($p_{t,L}/p_{t,R}$).
Those values are checked by
analytic solutions \citep{giac06}.
For example, in the reference cases ($p_{t,L}/p_{t,R}=10$),
the theory predicts $\gamma_{BL}/\gamma_{jet} \lesssim 1.78$
(Equations \ref{eq:max} and \ref{eq:max_para}) 
and $\gamma_{BL}/\gamma_{jet} \lesssim 3.16$
(Equation \ref{eq:max_perp}),
while we obtain $\gamma_{BL}/\gamma_{jet}=1.67$ (H1), $1.59$ (M1), and $2.31$ (M2)
(see also Figure \ref{fig:profile}{\itshape c}).
In general, one can see that
the scaling laws are in excellent agreement with
the boost amplitude in the H1 and M1 series.
The M1 cases are slightly affected by
another limitation (e.g., Equation \ref{eq:max2}),
due to the lower initial temperature $(p_{g,L}/\rho_L)$ in the jet. 
In the case of the M2 series,
Equation \ref{eq:max_perp} works as a looser upper limit.
Since the theory is valid
when the magnetic pressure dominates in the jet, $p_{t,L}\sim b_L^2/2$,
it is reasonable that we obtain intermediate results
in these specific cases. 

We perform another parameter survey
by reducing the jet-side density $\rho_{L}$ (Table \ref{table}).
The results are very similar.
Since $p_{L} \gg \rho_{L}$, we have
even better agreement with the theory in the M1 series.

\section{Relevance for kinetic models}

In this section, we examine the problem
from the viewpoint of the kinetic theory.
For brevity, we assume that
the gas moves to the $+z$-direction with a speed of $\beta=v_z$,
and we set the particle rest mass to $m=1$.
Although the RHD theory does not assume
a specific distribution function,
a drifting Maxwellian \citep{jut11,synge}
will be the best starting point:
\begin{eqnarray}
\label{eq:js}
f (\vec{p}) d\vec{p}
~ \propto ~
\exp\Big[ -\frac{ \gamma (p_0 - \beta p_z) }{ T } \Big]
~d\vec{p}
,
\end{eqnarray}
where $\vec{p}$ is the particle momentum,
$p_0=[{1+(\vec{p}\cdot\vec{p})}]^{1/2}$ is the particle energy, and
$\gamma,\beta$ are the fluid bulk properties.

\begin{figure}[thbp]
\begin{center}
\ifjournal
\includegraphics[width={\columnwidth},clip]{f4.eps}
\else
\includegraphics[width={\columnwidth},clip]{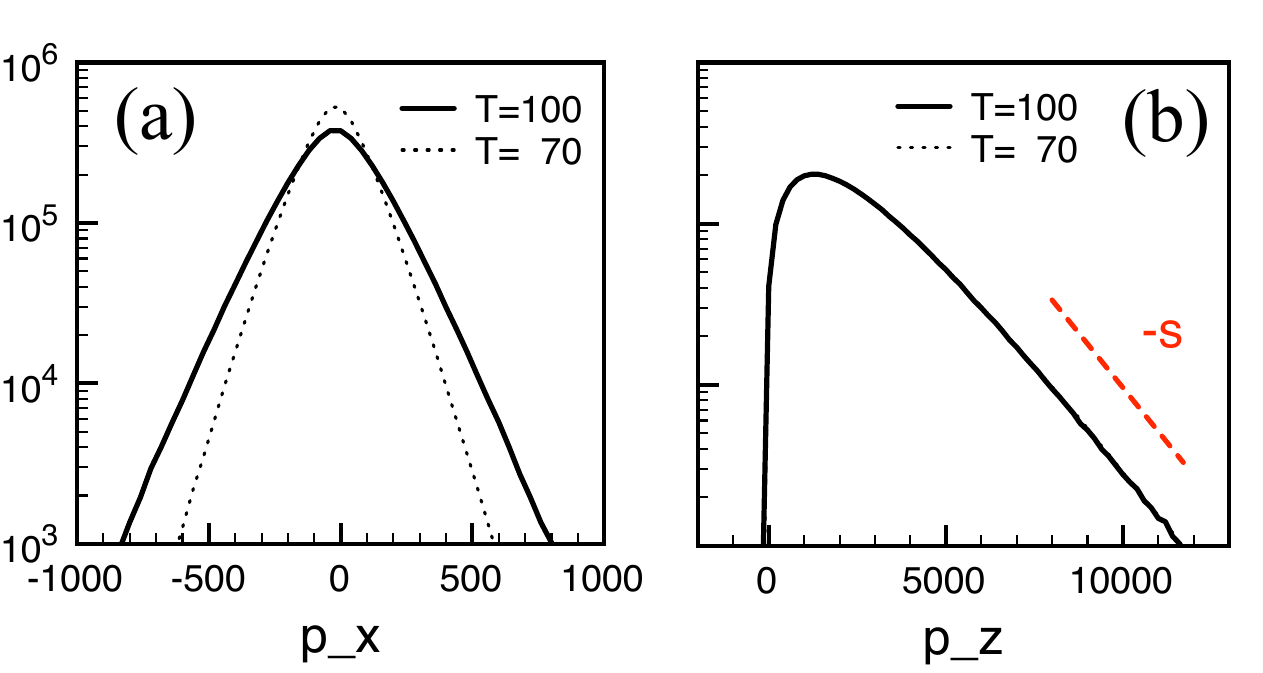}
\fi
\caption{
(Color online)
(a) Ideal gas distribution functions for
(1) $T=100$ and $\gamma=7$ ({\itshape solid line}) and
(2) $T=70$ and $\gamma=10$ ({\itshape dotted line})
in the $p_x$-space.
(b) The same, but in the $p_z$-space.
\label{fig:PDF}}
\end{center}
\end{figure}

Shown in Figure \ref{fig:PDF} are
momentum-space profiles of sample distribution functions.
Two samples are generated by Equation \ref{eq:js}:
(1) $T=100$ and $\gamma=7$ and (2) $T=70$ and $\gamma=10$
such that they satisfy Equation \ref{eq:rT}. 
We intend to mimic 
(1) the initial condition in the jet and
(2) the evolved population in the rarefaction region,
in model H1.
The lab-frame density $\gamma \rho$ is set to the same.

The $p_x$-profiles (Figure \ref{fig:PDF}{\itshape a})
are reasonably different due to the thermal spread.
In contrast, the $p_z$-profiles (Figure \ref{fig:PDF}{\itshape b}),
which significantly extend to the $+p_z$ direction, look quite similar. 
In the left side of the $p_z$-space ($p_z \ll 0$),
from Equation \ref{eq:js} and $p_z \approx -p_0$,
the asymptotic slope index $s$ of the distribution
$F(p_z) \propto e^{s p_z}$ yields 
\begin{eqnarray}
s \sim \frac{\gamma (1+\beta)}{T}
\sim \frac{2\gamma}{T}.
\end{eqnarray}
We see that the population is quite limited in this side,
when $\gamma$ is large.
In the right side, the index $s$ will be
\begin{eqnarray}
\label{eq:s}
s \sim -\frac{\gamma (1 - \beta)}{T}
\sim -\frac{1}{2\gamma T} 
\sim {\rm const.}
\end{eqnarray}
Therefore the $p_z$-profile remains similar in this side,
even when the ``fluid'' velocity changes. 
In addition, since the right-side population
mainly carries the momentum and the energy,
the two distributions carry nearly the same amount of
the momentum and the energy density
per the lab-frame density,
as mentioned by Equations \ref{eq:constP} and \ref{eq:ene2}. 
The relative differences are
0.3\% in momentum and 0.6\% in energy, respectively.

Important implication of Equation \ref{eq:s} is
that the typical momentum spread is
$s^{-1} \sim 2\gamma T$ in the $+p_z$-direction.
Recalling the effective boost condition of $T \gg 1$,
we see that
a thermal umbrella is much bigger ($2T$ times)
than the bulk Lorentz factor $\gamma$
in the relativistic momentum space. 


\section{Discussion and Summary}

As shown in Equation \ref{eq:dpdt},
the anomalous bulk boost comes from
the temporal decrease of relativistic pressure.
From the energy viewpoint, the term transports
the internal energy to that of the bulk motion ($p_g \Rightarrow \gamma$),
as mentioned by \citet{aloy08}.
The internal-to-bulk energy transport is somewhat counter-intuitive,
however, it is a logical consequence of the relativistic fluid formalism.

The site of the boost is the rarefaction region.
The rarefaction wave involves the temporal pressure decrease behind its wave front
and there is a room for the convective fluid motion (Equation \ref{eq:p_force}).
In contrast, neither conditions are satisfied around the shocks.
The anomalous boost does not occur
on the other side of the contact/tangential discontinuity nor
will it occur when another shock replaces the rarefaction wave.
Therefore, the transition from the shock regime to
the rarefaction wave regime \citep{rez02}
would be a critical condition for the problem. 
Similar boost in the normal direction
has recently been reported
in magnetically-dominated rarefaction region as well
\citep{mizuno09}.

Another explanation is
a relativistic free expansion in the jet frame \citep{kom09b}.
When the relativistically strong pressure pushes
the gas outward against the external medium,
the lateral expansion can be relativistic in the jet frame.
Then, the Lorentz factor in the observer frame yields
$\gamma_{BL} \sim \gamma_{jet} (1-v'^2)^{-1/2}$,
where $v'$ is the expansion speed in the jet frame.
We expect that the term $-\vec{v}'\partial_{t'} p_t$ enhances
such expansion in the rarefaction region, and that
the relevant boost is projected into the tangential boost in the observer frame.
Strictly speaking, a 1D problem in the observer frame is no longer
identical to that in the jet frame,
because a 1D expansion of the discontinuity front
in the +$x$-direction is projected to
the oblique direction in the jet frame.
The two problems start differently
and therefore the situation is more complicated.

A potential limitation is that
multi-dimensional instabilities may modulate the 1D evolution.
Especially, the relativistic Kelvin--Helmholtz (KH) instabilities
will be relevant.  In the regime of our interest,
the increasing Lorentz factor \citep{tur76,bp76,bodo04} and
the flow-aligned magnetic field \citep{osm08}
suppress the KH mode;
for instance, if we employ
\citet{bodo04}'s stability condition of
$\gamma_{jet} > ( 1+2 \cos^{-2}{\theta} )$ in our RHD jet ($\gamma_{jet}=7$),
where $\theta$ is the angle between the jet flow and the wavevector,
the instability is allowed only in the quasi-transverse direction.
On the other hand, shear layers with density asymmetry
are known to be substantially KH-unstable. 
Once the KH vortex develops, the subsequent turbulence
is likely to smooth the sharp lateral structure. 
While 1D-like signatures have been found
in some three-dimensional RHD \citep{aloy05} and two-dimensional RMHD simulations
\citep{mizuno08,tch09,kom09b},
interference with the KH and other instabilities
needs further investigation. 

In addition, we need to keep in mind that
the entire process depends on
the ideal fluid assumption.
In order to justify it,
collisional or other scattering processes
have to relax the gas
much quicker than the dynamical timescale. 
However, those are difficult conditions especially in the jet side,
where the physical processes look even slower by the relativistic effect. 
In Section 5, we show that the fluid bulk speed is
considerably smaller than a wide thermal spread
in the momentum profile, when the boost operates. 
We think that
the counter-intuitive force may be just enforced
by the ideal fluid assumption:
i.e. the anomalous fluid acceleration may be
an artifact of an expedient isotropic fluid velocity. 
In the real world,
we expect that non-ideal effects such as the heat flow play roles. 
In fact, the system involves large gradient of the pressure and the temperature
in the rarefaction regions and around the discontinuities. 
In the high-temperature regime of $T \gg 1$,
the energy and momentum balances are mainly controlled by
the pressure parts (the internal energy or the enthalpy flux),
which can be sensitive to the local gas distribution functions. 

In summary, we examined the 1D anomalous relativistic boost
\citep{aloy06,mizuno08} at the lateral boundary of relativistic jets.
We numerically and theoretically confirmed that
the anomalous boost occurs in the RHD and RMHD regimes.
We further derived simple scaling laws for the accelerated Lorentz factor,
\begin{eqnarray}
\gamma_{BL} \lesssim \gamma_{jet} \Big(\frac{p_{t,L}}{p_{t,R}}\Big)^s
\left\{
\begin{array}{cl}
s=1/4 & ~~({\rm hydro,~parallel})\\
s=1/2 & ~~({\rm perpendicular}) \\
\end{array}
\right.
\nonumber
\end{eqnarray}
We also note that the process operates in an {\itshape ideal} fluid.
The non-ideal effects (heat flow etc.)
as well as multi-dimensional effects
are left for future works.
We hope that this work will be
a basic piece for the boundary problems
in relativistic jets and the relevant simulations.

\begin{acknowledgments}
The authors express their gratitude to Tadas Nakamura,
Karl Schindler, Yosuke Matsumoto, and Masha Kuznetsova for helpful comments. 
S.Z. gratefully acknowledges support from NASA Postdoctoral Program.
\end{acknowledgments}

\ifjournal
\clearpage 
\fi


\begin{thebibliography}{}
\bibitem[Aloy et al.(2005)]{aloy05} Aloy, M. A., Janka, H.-T., \& M{\"u}ller, E. 2005, \aap, 436, 273
\bibitem[Aloy \& Mimica(2008)]{aloy08} Aloy, M. A., \& Mimica, P. 2008, \apj, 681, 84
\bibitem[Aloy \& Rezzolla(2006)]{aloy06} Aloy, M. A., \& Rezzolla, L. 2006, \apj, 640, L115
\bibitem[Anile(1989)]{anile89} {Anile, A. M.} 1989, ``{Relativistic Fluids and Magneto-fluids},'' {Cambridge Univ. Press}
\bibitem[Birkinshaw(1991)]{birk91} Birkinshaw, M. 1991, in ``Beams and Jets in Astrophysics'' ed. Hughes, P.~A. (Cambridge Univ. Press), 278
\bibitem[Blandford \& Pringle(1976)]{bp76} Blandford, R. D., \& Pringle, J. E. 1976, \mnras, 176, 443
\bibitem[Bodo et al.(2004)]{bodo04} Bodo, G., Mignone, A., \& Rosner, R. 2004, \pre, 70, 6304
\bibitem[Ferrari(1998)]{ferrari98} Ferrari, A. 1998, \araa, 36, 539
\bibitem[Ferrari et al.(1980)]{ferrari80} Ferrari, A., Trussoni, E., \& Zaninetti, L. 1980, \mnras, 193, 469
\bibitem[Giacomazzo \& Rezzolla(2006)]{giac06} Giacomazzo, B., \& Rezzolla, L. 2006, {\itshape J. Fluid Mech.}, 562, 223
\bibitem[J{\"u}ttner(1911)]{jut11} J{\"u}ttner, F. 1911, {\itshape Ann. Phys.}, 339, 856
\bibitem[Komissarov et al.(2009)]{kom09b} Komissarov, S.~S., Vlahakis, N., \& K\"{o}nigl, A. 2009, \mnras, {\itshape submitted} (arXiv:0912.0845)
\bibitem[M{\'e}sz{\'a}ros(2006)]{mes06} M{\'e}sz{\'a}ros, P. 2006, {\itshape Rep. Prog. Phys.}, 69, 2259
\bibitem[Mignone \& McKinney(2007)]{mig07b} Mignone, A., \& McKinney, J. C., 2007, \mnras, 378, 1118
\bibitem[Mignone et al.(2009)]{mig09} Mignone, A., Ugliano, M., \& Bodo, G. 2009, \mnras, 393, 1141
\bibitem[Mirabel \& Rodr\'{i}guez(1999)]{mirabel99}
Mirabel, I. F., \& Rodr\'{i}guez, L. F. 1999, \araa, 37, 409
\bibitem[Miyoshi \& Kusano(2005)]{miyoshi05} Miyoshi, T., \& Kusano, K. 2005, \jcp, 208, 315
\bibitem[Mizuno et al.(2008)]{mizuno08} Mizuno, Y., Hardee, P., Hartmann, D. H., Nishikawa, K.-I., \& Zhang, B. 2008, \apj, 672, 72
\bibitem[Mizuno et al.(2009)]{mizuno09} Mizuno, Y., Zhang, B., Giancomazzo, B., Nishikawa, K.-I., Hardee, P., Nagataki, S., \& Hartmann, D. H. 2009, \apj, 690, L47
\bibitem[Osmanov et al.(2008)]{osm08} Osmanov, Z., Mignone, A., Massaglia, S., Bodo, G., \& Ferrari, A.  2008, \aap, 490, 493
\bibitem[Ostrowski(2000)]{ost00} Ostrowski, M 2000, \mnras, 312, 579
\bibitem[Piran(2004)]{piran04} Piran, T. 2004, {\itshape Rev. Mod. Phys}, 76, 1143
\bibitem[Rezzolla \& Zanotti(2002)]{rez02} Rezzolla, L., \& Zanotti, O. 2002, \prl, 89, 114501
\bibitem[Romero et al.(2005)]{romero05} Romero, R., Mart{\'{\i}}, J.~M., Pons, J.~A., Ib{\'a}{\~n}ez, J.~M. \& Miralles, J.~A. 2005, {J. Fluid Mech.}, 544, 323
\bibitem[Sakai \& Kawata(1980)]{sakai} Sakai, J., \& Kawata, T. 1980, {J. Phys. Soc. Japan}, 49, 747
\bibitem[Stawarz \& Ostrowski(2002)]{so02} Stawarz, {\L}., \& Ostrowski, M. 2002, \apj, 578, 763
\bibitem[Synge(1957)]{synge} Synge, J. L. 1957, The Relativistic Gas (New York: Interscience)
\bibitem[Tchekhovskoy et al.(2009)]{tch09} Tchekhovskoy, A., Narayan, R., \& McKinney, J.~C. 2009, \apj, {\itshape submitted} (arXiv:0909.0011)
\bibitem[Turland \& Scheuer(1976)]{tur76} Turland, B. D., \& Scheuer, P. A. G. 1976, \mnras, 176, 421
\bibitem[Urry \& Padovani(1995)]{up95} Urry, C.~M.\& Padovani, P. 1995, PASP, 107, 803
\bibitem[van Leer(1977)]{mc} van Leer, B. 1977, \jcp, 23, 276
\end{thebibliography}
\end{document}